\documentclass{aa}

\usepackage{graphicx}
\usepackage{natbib}

\bibpunct{(}{)}{;}{a}{}{,}
\usepackage{txfonts}

\begin{document}
\title{Modeling the gas-phase chemistry of the transitional disk around HD~141569A}

\author{B. Jonkheid\inst{1}\and I. Kamp\inst{2}\and J.-C. Augereau\inst{1,3}\and E.F. van Dishoeck\inst{1}}

\authorrunning{Jonkheid et al.}
\titlerunning{Modeling the gas around HD~141569A}

\offprints{B. Jonkheid, \email{jonkheid@strw.leidenuniv.nl}}

\institute{
Sterrewacht Leiden, P.O. Box 9513, 2300 RA Leiden, the Netherlands
\and Space Telescope Division of ESA, Space Telescope Science Institute, 3700 San Martin drive, Baltimore, MD 21218, USA
\and Laboratoire d'Astrophysique, Observatoire de Grenoble, BP 53, F-380431 Grenoble C\'edex 9, France}

\date{Received 27 December 2005 / Accepted 14 March 2006}

\abstract{}{The chemistry, distribution and mass of the gas in the transitional disk around the 5 Myr old B9.5 V star HD~141569A are constrained.}{A quasi 2-dimensional (2D) chemistry code for photon dominated regions (PDR) is used to calculate the chemistry and gas temperatures in the disk. The calculations are performed for several gas distributions, PAH abundances and values of the total gas mass. The resulting CO $J=2-1$ and $J=3-2$ emission lines are computed with a 2D radiative transfer code and are compared to observations.}{The CO abundance is very sensitive to the total disk mass because the disk is in a regime where self-shielding just sets in. The observed CO emission lines are best fit by a power-law gas distribution of $80\,M_\oplus$ starting at 80 AU from the central star, indicating that there is some gas in the inner hole. Predictions are made for intensities of atomic fine-structure lines. [\ion{C}{i}], which is the dominant form of carbon in large parts of the disk, is found to be a good alternative tracer of the gas mass.}{}


\maketitle

\section{Introduction}

Circumstellar disks are the natural by-product of star formation, and once a
protostar has formed they are crucial for the further accretion of matter onto 
the star \citep{shu87}. Over the course of its life a circumstellar disk 
loses its gas until a dusty debris disk remains in which the small dust grains 
are produced by collisions of planetesimals. The transformation from 
gas-rich accretion disks to dusty debris disks occurs relatively quickly 
\citep{simpra95}, and as a result only a few objects in this transitional stage
have been found. One example is the disk around HD~141569A, which is a
pre-main sequence Herbig Ae/Be star with spectral type B9.5 V at a distance of 
99 pc. It has a luminosity of 22 $L_\odot$ \citep{merin04}. Its age is 
estimated at $\sim 5\pm 3\,{\rm Myr}$ \citep{weinbe00,merin04} and it has two 
M-type companions HD~141569B and C \citep{rossit43}. The optically thin dust 
disk around HD~141569A has 
been observed in infrared emission \citep[e.g.][]{sylves96} and in scattered 
light at near infrared \citep{augere99,weinbe99} and visible wavelengths
\citep{mouill01,clampi03}. Evidence for both large dust grains \citep{boccal03}
and polycyclic aromatic hydrocarbons \citep[PAHs, see][]{sylves96,weinbe04} has
been found. Gas in the disk has also
been detected in the form of CO pure rotational lines \citep{zucker95,dent05} 
and ro-vibrational lines \citep{britta03}. Thus, this disk provides a unique
laboratory to study the effects of dust evolution on the gas.

Observations of the dust in the HD~141569A disk show a complex morphology:
there is a large inner hole in the dust distribution out to 150 AU from the 
central star and two dust rings at 185 AU and 325 AU 
\citep{augere99,weinbe99,mouill01,boccal03,clampi03}. The outer dust ring may 
be explained by tidal interactions of the disk with the M-type companions 
\citep{augpap04}, 
while both rings may also have formed due to hydrodynamic drag forces on 
the dust grains \citep{klalin05} or interaction with a giant planet 
\citep{wyatt05}. The origin of the large inner hole is still unclear but it 
could contain PAHs (Weinberger et al., in preparation).

\begin{figure*}[!tp]
\centering
\includegraphics[width=17cm,height=13.6cm]{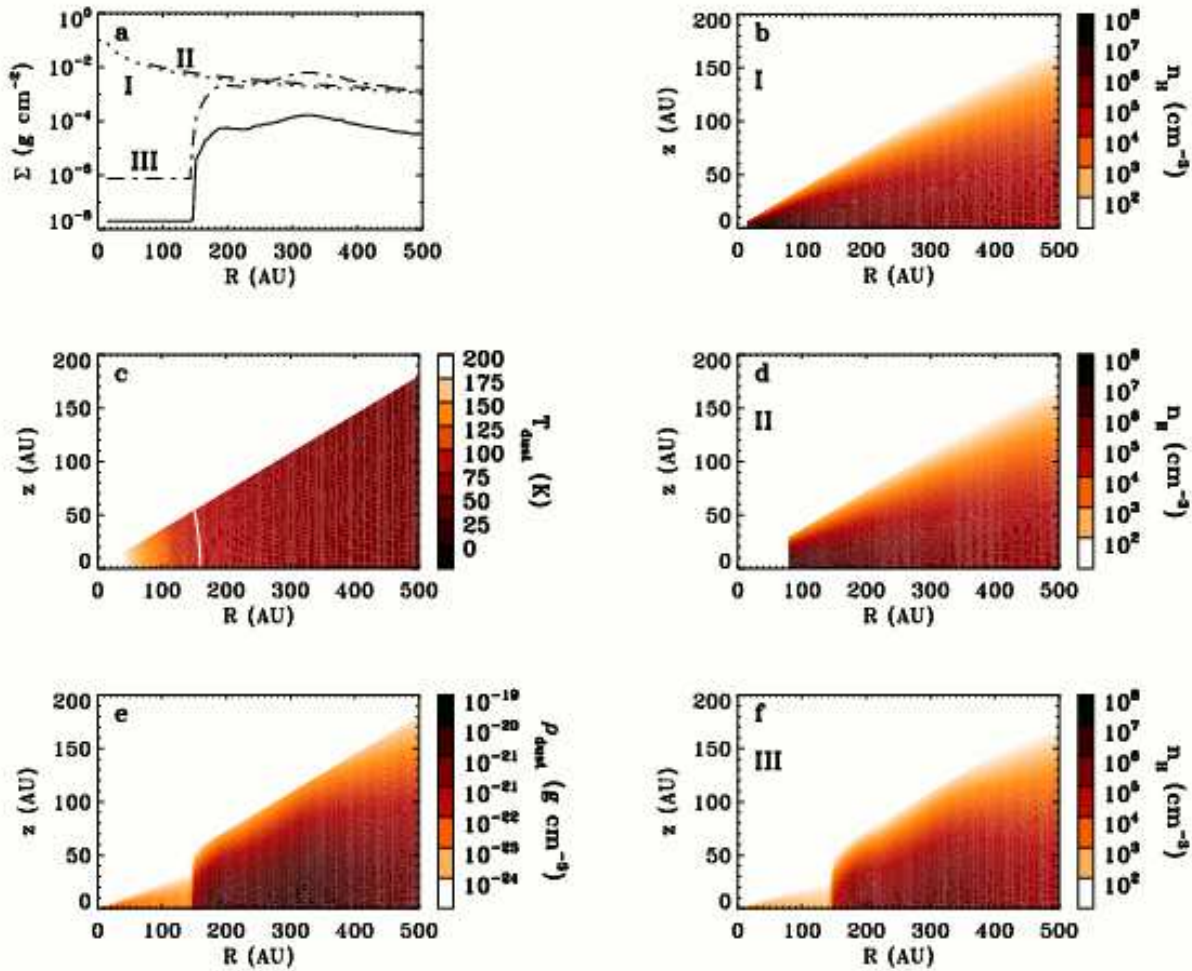}
\caption{Input parameters of the disk model. Panel a gives the surface densities for dust (solid line) and for three different distributions of gas (dashed lines). Panels b, d and f give the corresponding gas number densities. Panel c gives the mean dust temperature (averaged over the grain size distribution), where the white line denotes the 100 K isotherm. Panel e gives the mass density of dust grains.}
\label{struct}
\end{figure*}

The single dish CO observations by \citet{zucker95} and \citet{dent05}
in principle provide constraints on the mass of gas in the 
disk. Since this object is in a transitional stage it should be losing its gas 
rapidly, and determining the gas mass and gas/dust ratio may give insight into 
the processes responsible for the gas loss. Additionally, the total amount of 
gas in the disk
has implications on the dynamics of the dust grains, and an estimate of the gas
mass and its distribution may help future dynamical simulations of the disk.
In this paper a chemical model is presented which calculates molecular 
abundances at each position in the disk as well as line intensities to compare 
with submillimeter observations, building on the work by \citet{kamzad01}, 
\citet{kamp03} and \citet{jonkhe04}. By varying the gas distribution and the 
total gas mass of the disk, a best fit to the CO observations is presented.
Predictions for other gas tracers are made.

\section{Model}

\subsection{Input}

\begin{table}
\centering
\caption[]{Adopted gas-phase elemental abundances with respect to hydrogen.}
\label{abuntab}
\begin{tabular}{l l}
\hline
Element & abundance\\
\hline
C & $1.3\times 10^{-4}$\\
O & $2.9\times 10^{-4}$\\
Mg & $4.2\times 10^{-6}$\\
S & $1.9\times 10^{-6}$\\
Si & $8.0\times 10^{-6}$\\
Fe & $4.3\times 10^{-6}$\\
PAH & $ 1-2\times 10^{-10\ {\rm a}}$\\
\hline
\end{tabular}
\begin{enumerate}
\item[${\rm ^a}$] The PAH abundance varies between models with different gas masses, since the total mass in PAHs is kept constant.
\end{enumerate}
\end{table}

The basis for the calculations presented here is formed by the 1+1-dimensional 
model described in \citet{jonkhe04}. In this model the disk is divided into a 
series of vertical 1-dimensional structures, each of which was treated as a 
photon dominated region (PDR) illuminated from above. The PDR code 
calculates the chemistry using the detailed radiative transfer of ${\rm H_2}$ 
and CO dissociating lines described by \citet{bladis87} and \citet{disbla88}.
The elemental abundances used in the calculations are displayed in Table 
\ref{abuntab}; it is assumed that the gas in the disk is of interstellar origin
instead of second generation gas such as may result from the evaporation of
solid bodies.
The thermal balance is solved taking into account thermal coupling between gas 
and dust, heating through photoionization of polycyclic hydrocarbons (PAHs),  
photoelectric effect on large grains, and the formation and dissociation of 
${\rm H_2}$ and 
cooling through CO rotational lines and the fine-structure lines of C, 
${\rm C^+}$ and O. The code was tested extensively against other codes in a 
benchmark project (Roellig et al., in preparation).

\begin{figure}
\resizebox{\hsize}{!}{\includegraphics[angle=0]{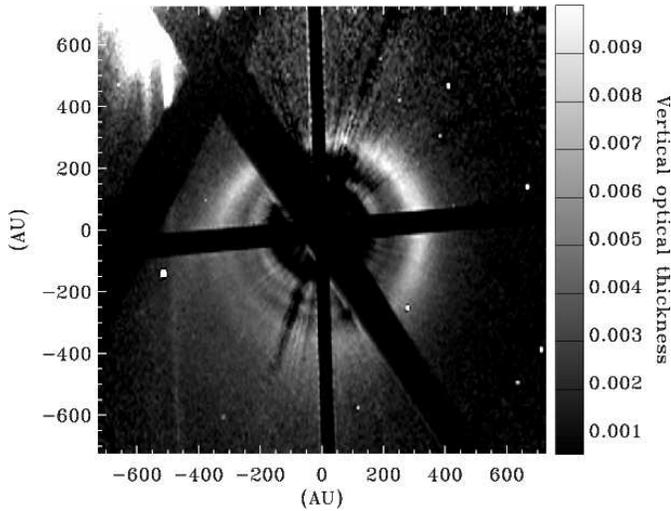}}
\caption{Deprojected HST image of the HD 141569 disk in scattered light by \citet{mouill01}.}
\label{hddust}
\end{figure}

Since the HD~141569A disk differs significantly in terms of mass and structure 
from the T-Tauri disks examined in \citet{jonkhe04}, a different input is used
(see Figure \ref{struct}). The dust distribution is taken from \citet{mouill01}
by deprojecting the disk assuming a scattering phase function with $g=0.2$ in 
the V-band (see Figure \ref{hddust}). The dust content in the inner hole is 
uncertain: the upper limit to the surface density is 
$2\times10^{-6}\,{\rm g\,cm^{-2}}$. In the 
calculations a value of $2\times10^{-8}\,{\rm g\,cm^{-2}}$ is used at $R<150$ 
AU to be on the conservative side. An azimuthal average of the distribution was
then taken to obtain the surface density in Figure \ref{struct}a. 

Several trial profiles were used for the radial gas 
distribution: a power law similar to the minimal mass solar nebula where 
$\Sigma_{\rm gas}\propto R^{-1.2}$, with an inner radius of either 15 AU 
(called distribution I, Figure \ref{struct}b) or 80 AU (distribution II, Figure
\ref{struct}d), and a distribution similar to the dust distribution (with an 
effective inner radius of 150 AU, called distribution III, see Figure 
\ref{struct}f). 
For each assumption of the radial distribution the total 
gas mass was varied by scaling the entire distribution by a single factor.
The densities of gas and dust were derived from their surface densities 
assuming a vertical distribution of $n\propto e^{-z^2/2h^2}$, where the scale
 height
$h=0.085\, R$ and $R$ and $z$ are cylindrical coordinates. The densities shown 
in Figure \ref{struct}
for the different radial distributions are for total gas masses of 80 
$M_\oplus$. The dust temperature
is determined assuming the grains are in thermal balance and the disk is 
optically thin in continuum radiation at infrared through ultraviolet (UV) 
wavelengths. 

\begin{figure}
\resizebox{\hsize}{!}{\includegraphics[angle=0]{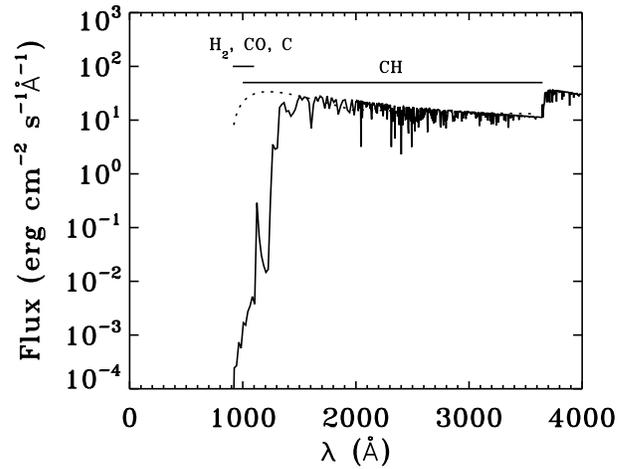}}
\caption{The flux at the inner edge of the gas disk in distribution I and II
(15 AU from the central star), for the radiation field of a B9.5 star (solid 
line) and of the interstellar radiation field (dotted line), scaled to fit the 
integrated flux between 912 and 3600 \AA, corresponding to $I_{\rm UV}=10^7$ 
times the interstellar radiation field. The spectral regime where ${\rm H_2}$ 
and CO are dissociated and C ionized, and the regime where CH is dissociated 
are indicated with bars at the top.}
\label{starspec}
\end{figure}

The interstellar 
spectrum used for the UV radiation by \citet{jonkhe04} is inappropriate 
for HD~141569 in the chemically important wavelength range of 
${\rm 912\,\AA<\lambda<1100\, \AA}$, so the spectrum of an A0 star 
\citep[taken from][with $T_{\rm eff}=10\,000$ K and $\log g = 4$, see Figure 
\ref{starspec}]{hausch99} was used here to calculate the photoionization and 
photodissociation rates. 
For the photoelectric heating rate the differences in the spectral shape of the
radiation field were implemented using the correction factor by 
\citet{spaans94} (Equation (13) in that paper). The radiation field has a 
strength of 
$I_{\rm UV}=10^7\times$ the strength of the interstellar radiation field by
\citet{draine78} at 15 AU from the central star; at 80 AU this factor is 
$3\times 10^5$.

\citet{jonkhe04} considered only more massive, optically thick disks, so a 
number of changes had to be made in the code to simulate this transitional 
object. First, the vertical PDR structures used in the 
original code are inappropriate here because scattering will be 
far less important in an optically thin disk. Therefore the PDRs are 
now taken radially from the star, with the strength of the radiation field 
scaling with $r^{-2}$ (where $r$ is the distance to the star, not to be 
confused with $R$, the radial component of the cylindrical coordinate system 
used in the plots) in addition to absorption effects.

\begin{figure*}[!tp]
\centering
\includegraphics[width=17cm,height=14.9cm]{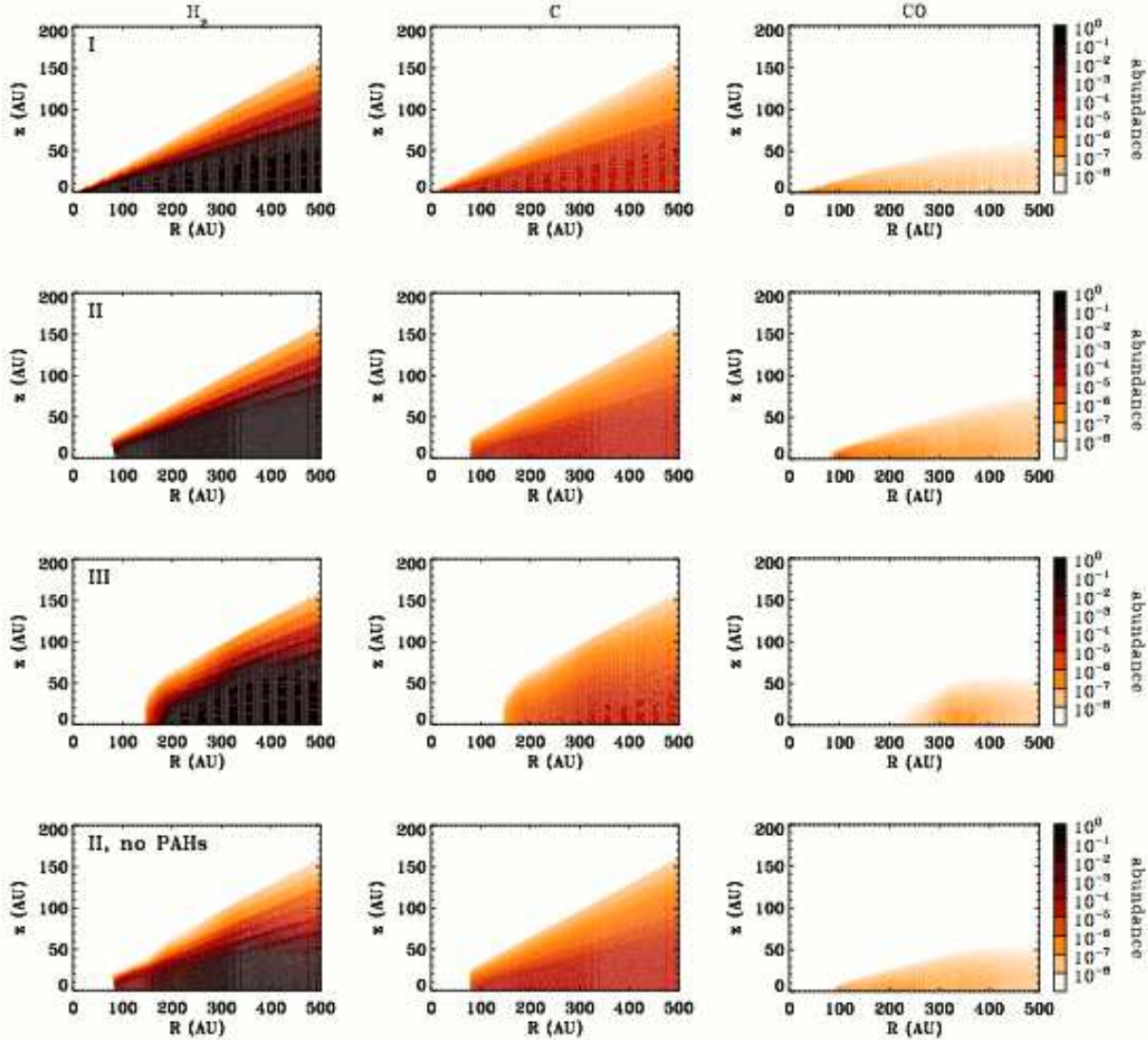}
\caption{Abundances (per H atom) for ${\rm H_2}$ (left column) C (middle column) and CO (right column)  for models I (first row), II (second row), III (third row) and II without PAHs (fourth row). In all cases the gas mass is 80 $M_\oplus$.}
\label{2dchem}
\end{figure*}

Second, the absorptions responsible for dissociation of ${\rm H_2}$ and CO and 
the ionization of C may
become optically thick if the column densities of these species toward the 
central star become sufficiently high. The interstellar 
radiation field then becomes the most important source of dissociation of these
molecules. This is implemented in the model by taking the photodissociation
rates for these molecules to be: 
$$R_{\rm ph,\,tot}=R_{\rm ph,\,star}+R_{\rm ph,\,ISM}$$
where $R_{\rm ph,\,tot}$ is the total photorate, 
$R_{\rm ph,\,star}$ is the photorate due to stellar light and 
$R_{\rm ph,\,ISM}$ is the photorate due to the interstellar 
radiation field. The former rates are calculated using the detailed treatments 
of \citet{bladis87} for ${\rm H_2}$,  \citet{disbla88} for CO and 
\citet{dishoe88} for C.
The latter rates are calculated assuming they have a value of 50\% of the 
unshielded interstellar photodissociation rate at
the disk's surface, with shielding calculated using the vertical column 
densities of ${\rm H_2}$, CO and C toward the nearest surface as inputs for 
the shielding functions by \citet{draber96} for ${\rm H_2}$, \citet{disbla88} 
for CO and \citet{werner70} for C. For all other
processes the interstellar radiation field was ignored, because the stellar 
radiation field has a strength of $I_{\rm UV}>1000\,\times$ 
the interstellar UV field of \citet{draine78} 
(disregarding absorption) even at 500 AU and dominates the continuum 
flux. 

\subsection{Treatment of dust}

Based on the scattering properties investigated by \citet{boccal03} and 
\citet{augpap04} and the SED
fitting by \citet{liluni03} the dust particles in the HD~141569A disk are found
 to be much larger than their counterparts in the interstellar medium: 
HD~141569 dust has grain sizes
${\rm 1\,\mu m} < a <{\rm 1\, cm}$ (with $a$ the equivalent spherical radius), 
and a size distribution $n(a)\propto a^{-3.3}$ \citep{liluni03}, while the ISM
grains have sizes ${\rm 5\,nm}< a <{\rm 0.25\,\mu m}$ and a distribution 
$n(a)\propto a^{-3.5}$ \citep{mathis77}. Since the 
Leiden PDR code assumes dust grains to have an interstellar size 
distribution, it has to be modified to incorporate the larger sized grains.
Dust grains enter the model in four ways: as absorbers for UV radiation, as
sites for ${\rm H_2}$ formation, and in the thermal balance as a source for the
photoelectric heating and gas-dust collisions. These processes are, to lowest 
order, proportional to either the surface area or the geometrical cross-section
of the grains, so they can be approximated by scaling them with 
$$\left<a^2\right>_{\rm HD~141569}/\left<a^2\right>_{\rm ISM}$$
where $<a^m>=\int_{a_{\rm min}}^{a_{\rm max}}a^m n(a)\,da/\int_{a_{\rm min}}^{a_{\rm max}} n(a)\,da$, and assuming the 
number of grains to remain constant in a given volume. If the mass in 
grains is kept constant rather than the number, all rates have to be mutiplied 
with a further factor of 
$$\frac{m_{\rm grain, ISM}}{m_{\rm grain, HD~141569}}=\frac{\left<a^3\right>_{\rm ISM}}{(1-P)\left<a^3\right>_{\rm HD~141569}}$$
where $P$ is the porosity of the grains in the HD~141569A disk. Following 
\citet{liluni03}, $P=0.73$ for ice-covered grains. Since most of the dust has
a temperature lower than 100 K, this value is used everywhere in the disk.
Thus, to lowest order all processes should 
be scaled by a factor 1/1500 using the grain parameters derived for HD~141569
for a gas/ dust mass ratio of 100.

This treatment of the dust properties leaves some inaccuracies of order unity,
the most important ones being in the extinction of stellar radiation and the 
photoelectric heating rate. The mean extinction cross sections and scattering 
properties were calculated independently using Mie theory with the dielectric 
function of \citet{liluni03} for porous icy aggregates (Equation (9) in that 
paper); the results were used to correct the extinction. The photoelectric 
heating rate was calculated using a photoelectric yield of $Y=0.05$ 
electrons/photon to simulate larger grains \citep{watson72}. For a single grain
size of $3\,{\rm \mu m}$ the resulting heating rates are similar to those used 
by \citet{kamzad01}; for the HD~141569 grains the appropriate size distribution
was used.

Even though the mean dust temperature shown in Figure \ref{struct} is high 
enough to prevent freeze-out of CO on the grains, there remains the possibility
that the large grains, which have a temperature lower than the mean dust 
temperature, act as sites for CO freezing. Calculations show that the
temperature of $> 100\,\mu$m sized grains falls below 30 K only in the outer 
regions of the disk, and that even millimeter sized grains have temperatures
$> 25$ K throughout the disk. It is therefore unlikely that freeze-out of 
CO will occur at any significant rate, so it is ignored in the calculations.

\begin{figure*}[!tp]
\centering
\includegraphics[width=17cm]{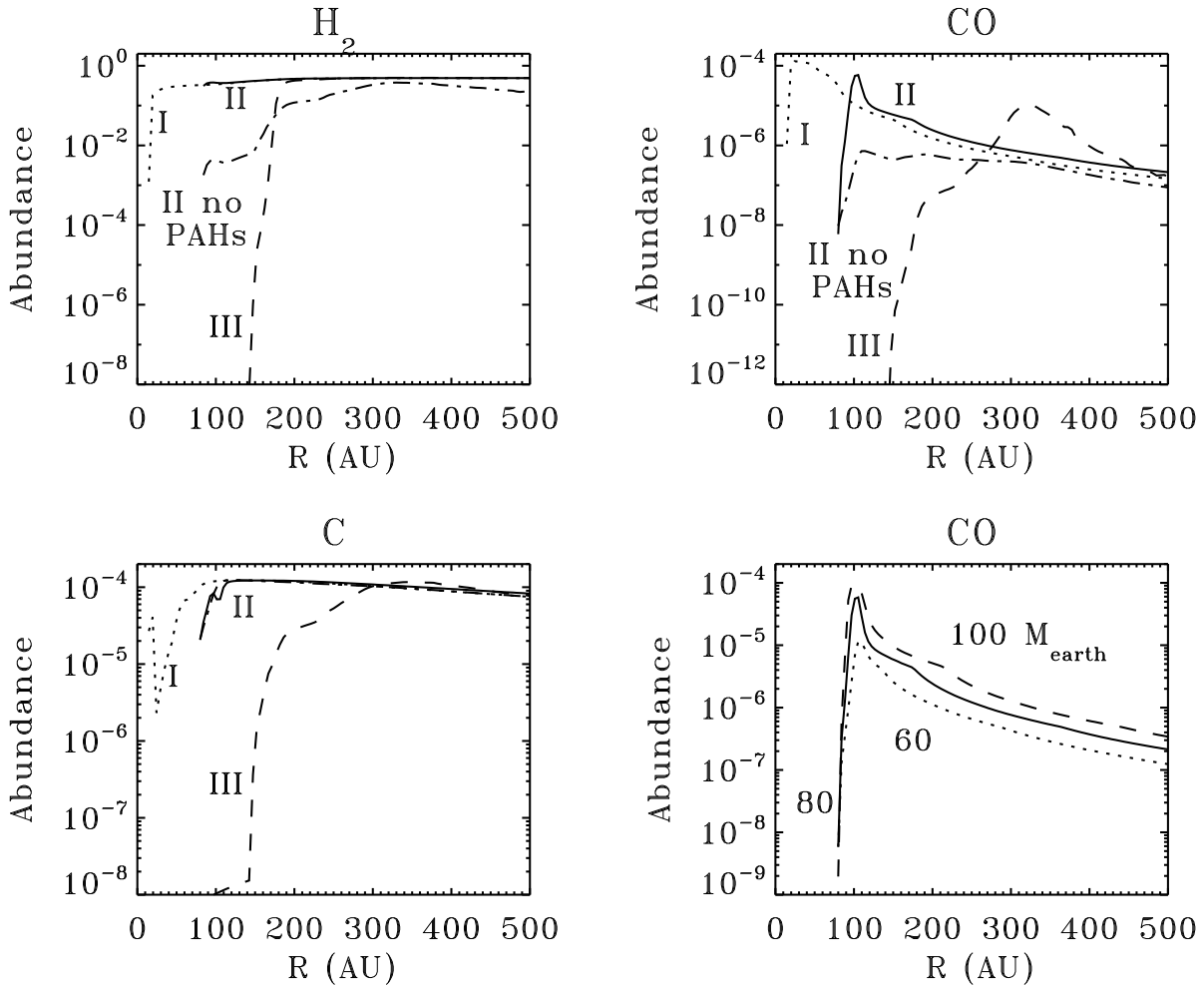}
\caption{Midplane abundances (per H atom) for ${\rm H_2}$ (upper left), CO (upper right) and C (lower left) for 80 $M_\oplus$ gas distributions I (dotted line), II (solid line) and III (dashed line), and for distribution II when PAHs are removed from the chemistry (dash-dotted line). The lower right panel gives the midplane abundance of CO for gas distribution II with total gas masses of 60 (dotted line), 80 (solid line) and 100 $M_\oplus$ (dashed line).}
\label{1dchem}
\end{figure*}

From infrared observations by \citet{sylves96} it is known that polycyclic 
hydrocarbons (PAHs) are present in the disk. In our model PAHs are assumed to 
be well mixed with the gas, i.e. the abundance of PAH 
molecules with respect to hydrogen is independent of location in the disk, even
in the inner hole. The total amount of PAHs is kept constant at 
$m_{\rm PAH}=7.9\times 10^{-6}\,M_\oplus$ \citep{liluni03}, regardless of the 
total gas mass. This means that the PAH abundance varies between models with 
different mass; their abundance is 
$x_{\rm PAH}=1.5\times 10^{-10}$ per H atom when 
$M_{\rm gas} = 80\,M_\oplus$. PAHs affect the disk's chemistry in three ways 
\citep[see][]{jonkhe04}: as sources of photoelectric heating, as absorbers of 
UV radiation and by direct chemical reactions (particularly in the formation of
${\rm H_2}$ and the charge balance). These processes are not only important in 
the inner disk where practically no large grains are present, but also in the 
outer disk due to the decreased efficiency of the large grains.

\begin{figure*}[!tp]
\centering
\includegraphics[width=17cm,height=17cm]{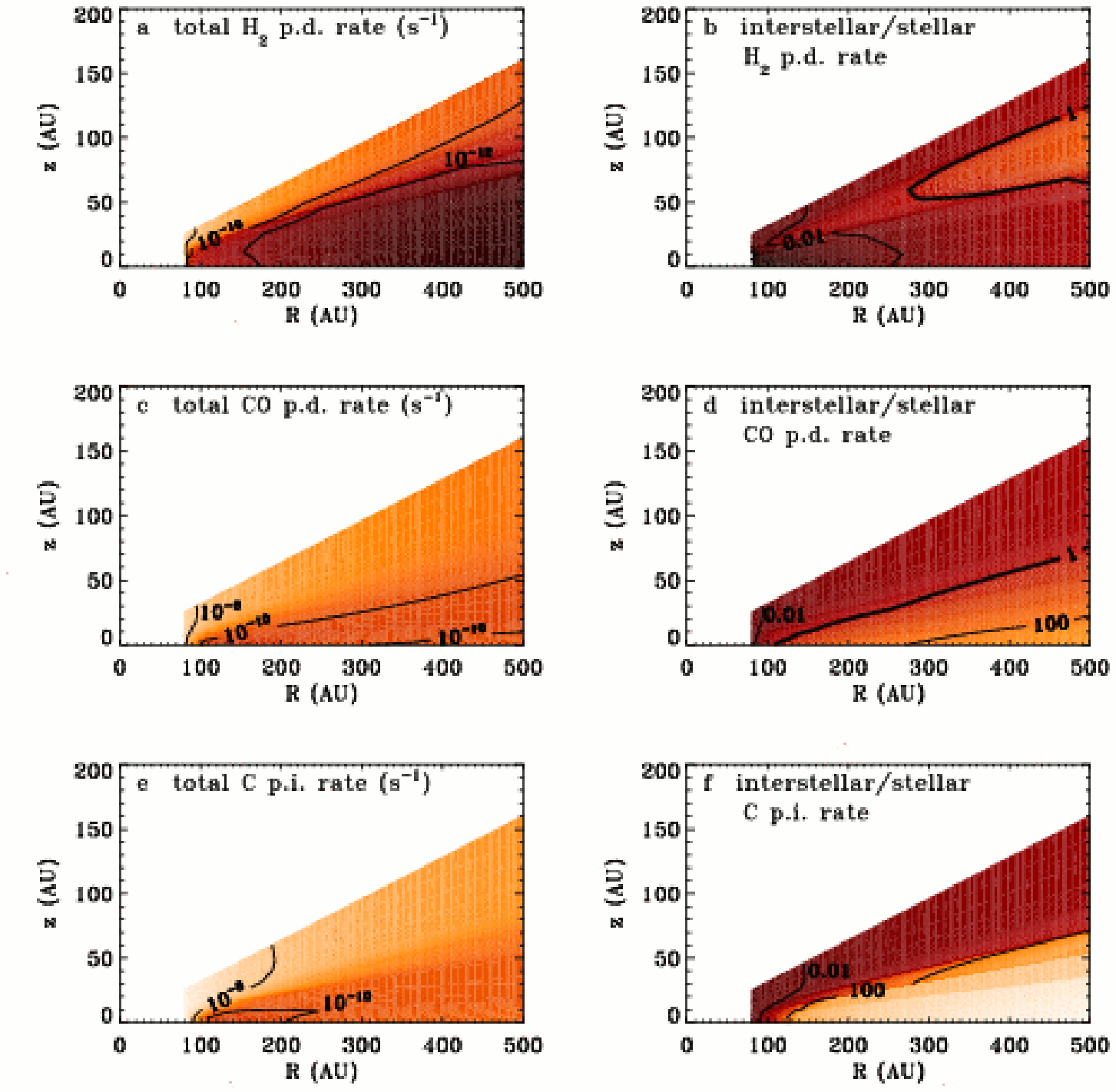}
\caption{The total photodissociation rates of ${\rm H_2}$ (panel a) and CO (panel c) and the total photoionization rate of C (panel e) in model II, and the ratio of the contibutions of the interstellar and stellar radiation fields (panels b, d and f). Results for models with different gas masses follow similar qualitative trends.}
\label{2ddiss}
\end{figure*}

\section{Results}

\begin{figure*}[!tp]
\centering
\includegraphics[width=17cm.height=9cm]{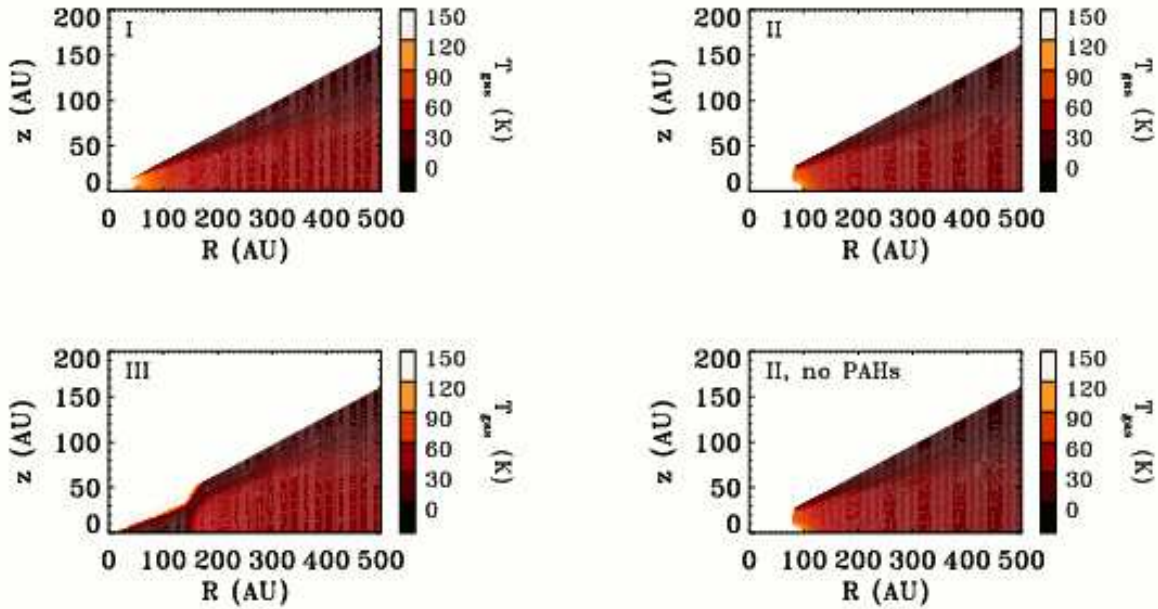}
\caption{The gas temperatures found for models I (upper left), II (upper right), III (lower left) and II without PAHs (lower right).}
\label{temps}
\end{figure*}

\begin{figure*}[!tp]
\centering
\includegraphics[width=17cm]{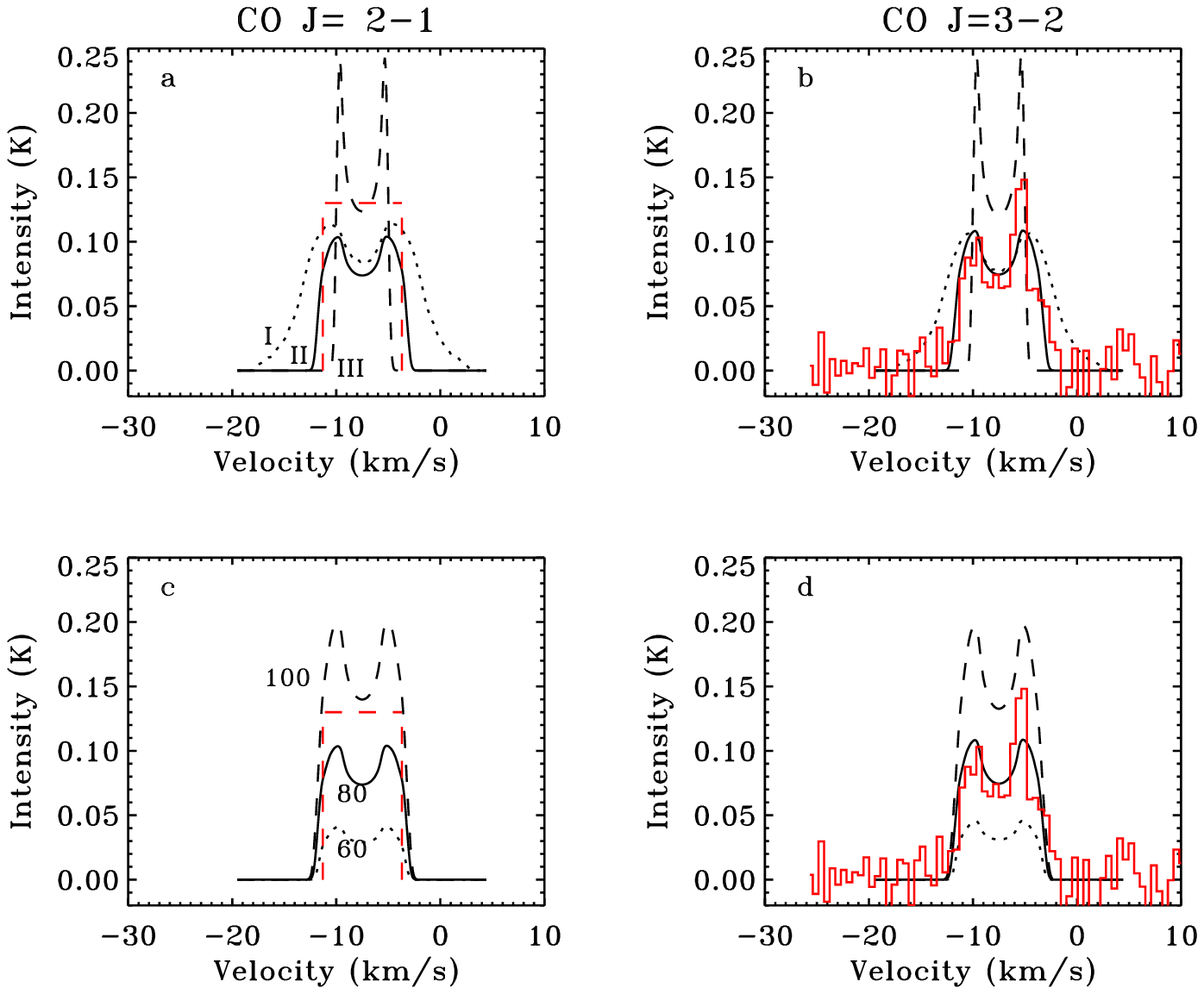}
\caption{CO $J=2-1$ (panels a and c) and $J=3-2$ (panels b and d) emission lines. Panels a and b give the best fit emission lines for the three different radial gas distributions shown in Figure \ref{struct}. The dotted line is for model I, the solid line model II, and the dashed line for model III. Panels c and d give the emission lines for model II with total gas masses of 60 $M_\oplus$ (dotted line), 80 $M_\oplus$ (solid line) and 100 $M_\oplus$ (dashed line). The red lines give the peak intensity and full width half maximum observed by \citet{zucker95} in the left panels, and the spectrum measured by \citet{dent05} for the right panels.}
\label{spect}
\end{figure*}

\subsection{Chemistry}
\label{chemsect}
The results for the chemistry for models with different radial gas 
distributions are shown in Figures \ref{2dchem} and \ref{1dchem}. 
It can be seen that models I and II show 
similar behaviour, with molecular abundances that are high at the inner edge of
the disk. Model III on the other hand produces ${\rm H_2}$ and CO only very far
in the outer regions of the disk. In all models the distribution of atomic 
carbon resembles the $\rm H_2$ distribution.

It can be seen in Figure \ref{1dchem} that in most models the carbon in the 
midplane is mostly atomic rather than ionized or molecular (in the form of CO).
The reason for this is that even though the radiation responsible for the 
photoionization of C and the photodissociation of CO is largely shielded 
by the upper 
layers, there are still many photons with $\lambda > 1200$ \AA\ left over
(see Figure \ref{starspec}). These photons
can proceed to dissociate precursors to CO (most notably CH), thereby 
decreasing the CO formation rate and thus driving 
the carbon chemistry to a neutral atomic phase.

In Figure \ref{1dchem} the CO abundances in the midplane are also shown for 
model II with total gas masses of 60, 80 and 100 $M_\oplus$. It can be seen 
that the CO abundance depends strongly on the gas mass, since it changes by
an order of magnitude with only a factor of 2 change in mass. This strong 
dependence on mass is due to self-shielding; in this regime shielding of CO
just sets in, and therefore every small increase in CO abundance in the upper 
layers enhances the production of CO in lower layers, producing a positive 
feedback loop.

Throughout the disk ${\rm H_2}$ is primarily formed on PAHs, and the large dust
grains are very inefficient absorbers of UV radiation. The chemical structure 
therefore shows no correlation with the dust distribution. If PAHs are removed 
from the chemistry ${\rm H_2}$ is primarily formed on large grains, and its
abundance resembles the dust distribution. Because of the decreased efficiency
of ${\rm H_2}$ formation on the large grains the molecular abundances are lower
than when PAHs are present. This is especially true for CO, which now cannot
build up sufficient column densities to become self-shielding.

It can be seen from 
Figure \ref{2ddiss} that the contribution of the interstellar UV field to the 
photodissociation of ${\rm H_2}$ and CO and the photoionization of C can be 
more than a factor of 100 higher 
than the stellar contribution in the outer disk, and therefore cannot be 
ignored in chemical calculations. This is especially true for C ionization, 
since the high abundance of atomic carbon in the inner disk shields the outer 
disk from the stellar radiation.
The abundance of ${\rm H_2}$ is high enough for this molecule to become 
completely self-shielding in the vertical direction as well as the radial 
direction. In the case of CO and C, the vertical self-shielding only just sets 
in. Thus the relative contribution of the interstellar UV field to ${\rm H_2}$ 
dissociation is not as large as it is for CO dissociation and C ionization.

The overall chemical timescales, which are controlled by the photorates, are 
short, ranging from from $<$ 100 yr in the inner disk and surface layers to a 
few $\times 10^3$ yr in the midplane of the outer disk. Thus chemical 
equilibrium is attained on dynamical timescales. 

\subsection{Temperature}

The results for the gas temperature are shown in Figure \ref{temps}. For models
I and II the surface layers of the disk are cool ($\sim 20\,{\rm K}$), because 
the photoelectric heating by PAHs and large grains is very inefficient here. In
the lower layers the abundances of C and ${\rm H_2}$ increase, and the gas is 
heated by the photoionization and photodissociation of these species, thus 
increasing the temperature with respect to the surface layers 
($\sim 50\, {\rm K}$). In model III the structure is slightly more complex: in 
the inner parts of the disk, the cooling is inefficient due to the low 
densities there. In the outer disk the heating by photodissociation of 
${\rm H_2}$ is very effective due to a combination of high abundances and high 
dissociation rate. It can be seen from Figure \ref{temps} that removal 
of the PAHs from the thermal balance has little effect on the temperature 
structure. The temperatures found in the disk's inner edge match the CO 
rotational temperature of 190 K found from the CO vibrational bands by 
\citet{britta03}. The temperatures found here and their distribution agree well
with the results of \citet{kamzad01} for the Vega disk. 

Initially no temperature solution could be found in the inner disk in some
models. It was found that the cooling rates for ${\rm C^+}$ and O were 
overestimated when the formulation of \citet{tiehol85} was used:
$$\Lambda_x(\nu_{ij})=n_i\,A_{ij}\,h\nu_{ij}\,\beta_{\rm esc}\,(\tau_{ij})\,C(\nu_{ij})$$
where $\Lambda_x$ is the cooling rate by species $x$, $n_i$ is the population
density of species $x$ in level $i$, $A$ is the spontaneous transition 
probability, $\beta_{\rm esc}$ is the escape probability and $C(\nu_{ij})$ is
a correction factor for the infrared background radiation from hot dust and the
cosmic microwave background radiation. In the Leiden PDR code UV pumping of the
fine structure levels of ${\rm C^+}$ and O is also included in calculating the 
populations, and the above formula becomes invalid for very high UV fields. In 
the inner disk, the UV pumping becomes the dominant process for populating 
these levels, and as a result the cooling rates were too large. This 
was solved by formulating the cooling rate as
$$\Lambda_x(\nu_{ij})=h\nu_{ij}(n_j k_{ji}-n_i k_{ij})$$
where $n_j$ and $n_i$ are the populations of the lower and upper levels, 
respectively, and $k_{ji}$ and $k_{ij}$ are the collisional excitation and
de-excitation rates. A stable temperature solution could be reached using this 
method. In the absence of UV pumping this reduces to the previous formulation.

\begin{figure*}[!tp]
\centering
\includegraphics[width=17cm]{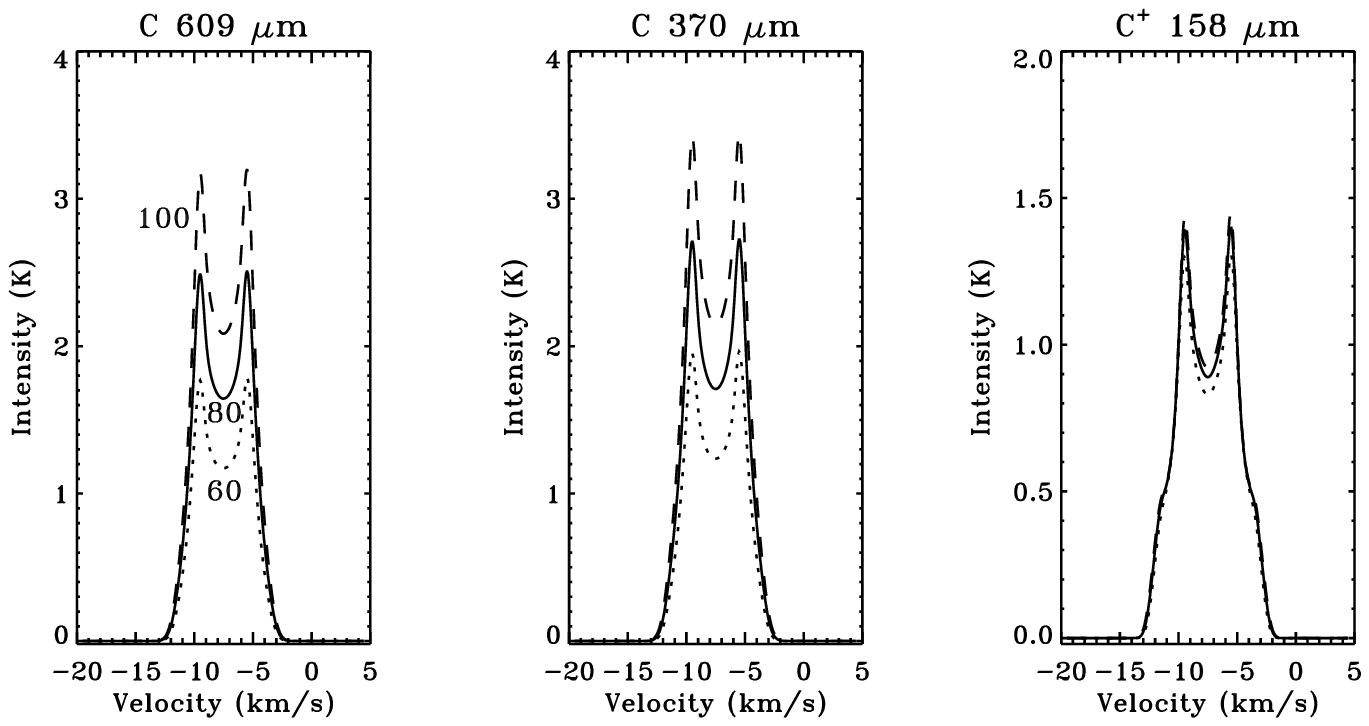}
\caption{The C and ${\rm C^+}$ fine structure lines for $M_{\rm gas}=60\,M_\oplus$ (dotted line), $80\,M_\oplus$ (solid line) and $100\,M_\oplus$ (dashed line).}
\label{cspec}
\end{figure*}

\section{Constraining the gas mass}
\label{lines}
\subsection{CO lines}
Figure \ref{spect} shows the predicted CO \mbox{$J=2\to 1$} and 
\mbox{$J=3\to 2$} 
submillimeter lines for 5 different models: the best fitting gas masses for the
radial gas distributions I, II and III (see Figure \ref{struct}), and two 
additional values of the gas mass for the preferred radial distribution II. 
The emission line profiles were calculated with the 2-dimensional radiative 
transfer code by \citet{hogtak00}. For these calculations, a distance to the 
HD~141569 system of 99 pc was used, and the disk is assumed to have an 
inclination of $55^\circ$ \citep{mouill01}. The results are convolved with a 
beam of 11 arcsec for the $2\to 1$ line (IRAM 30m at 230 GHz), and of 14 
arcsec for the $3\to 2$ line (JCMT 346 GHz). Of the former line only the peak 
intensity and FWHM are known; of the latter, the model results are compared 
directly to the observed spectrum. The computational grid used in the radiative
transfer code consists of 100 cells in the radial direction and 11 cells in the
vertical direction. 

It can be seen that the radial distribution of the gas mainly influences the 
widths of the lines investigated here. Distribution I gives line profiles that 
are too broad to fit the observations, while those of distribution III are too 
narrow. The intermediate distribution II fits the linewidths nicely.
Using this radial distribution the emission lines were calculated for 
several values of the total gas mass. The strong dependence of the CO abundance
on the gas mass shown in Figure \ref{2dchem} is reflected in the emission 
lines, where the peak intensity changes with a factor of 4 for less than a 
factor of 2 in mass. It can be seen that the 80 $M_\oplus$ model gives
a good fit to the CO $J=2\to 1$ and $J=3\to2$ emission lines.
The model predicts peak line intensities of 0.82 and 0.78 K for the
$J=4-3$ and $J=6-5$ lines in a 10$''$ beam.
The model cannot
account for the sharp peak seen in the red part of the $J=3-2$ line, however.
This peak is likely due to asymmetries in the disk structure, which can only be
investigated by a 3-dimensional disk model and a 3-D radiative transfer code.

There are several uncertainties in the derived gas mass of 80 $M_\oplus$: 
first, the elemental abundance of gas-phase carbon is 
uncertain by a factor of a few since the fraction of carbon locked in grains is
not known. Also, changes in other elemental abundances, and uncertainties in 
the PAH distribution and chemical rate coefficients affect the CO chemistry and
thus the intensity of the CO lines in relation to the gas mass.
Second, the CO chemistry is very sensitive to the stellar radiation field. If
the star has a higher flux in the 912-1100 \AA  range this would directly 
affect
the abundances of C and CO. Third, the assumption that ${\rm H_2}$ can form on 
PAHs is critical to the chemistry presented here. If the formation of 
${\rm H_2}$ is less efficient than assumed here it would reduce the abundance 
of all molecules in the disk, both because ${\rm H_2}$ is is a progenitor 
molecule for many species and because of the shielding to UV photons it 
supplies. For comparison
of results obtained with a completely independent chemical code 
\citep{kamzad01}, an uncertainty in the gas mass of a factor of a few is 
estimated.

The effect of disk inclination on the line profiles is limited. Spectra 
of the CO 3-2 line for distribution II with a gas mass of 80 $M_\oplus$ with
inclinations of $50^\circ$ and $60^\circ$ are very similar to the spectrum 
shown in Figure \ref{spect} and show equally good agreement with the observed 
spectrum. Since a $5^\circ$ change in inclination would indicate a 
morphological eccentricity of the disk of $\sim 0.45$, the value of $55^\circ$ 
by \citet{mouill01} is retained.

The models presented here confirm the presence of gas in the inner disk 
previously noted by \citet{zucker95}, \citet{britta03} and \citet{dent05}. 
Furthermore, the presence of vibrationally excited CO at temperatures of 
150-200 K  found by \citet{britta03} is matched by model II; model I has too
high temperatures in the inner disk, while model III only has CO in the outer 
disk where the radiation field is too low to excite the vibrational levels of 
CO and the temperatures are too low to match the observations. 

\subsection{Other emission lines}

The predicted [\ion{C}{i}] and [\ion{C}{ii}] fine structure lines at 609, 370 
and 158 $\mu$m are shown in Figure \ref{cspec}. The intensities were calculated
assuming radial distribution II, and assuming the beam is exactly the size of 
the disk (10 arcsec). Although the intensities of the [\ion{C}{i}] lines change
less dramatically with disk mass than the CO lines they are still a good 
--and probably more direct (see Figure \ref{2dchem})-- indicator of the total 
disk mass. The [\ion{C}{ii}] line on the other hand 
depends only very weakly on the disk mass; while there is more carbon in disks 
of higher mass, the fraction of ionized carbon is smaller.

It should be noted that the C abundance and [\ion{C}{i}] intensity are
sensitive to the UV intensity at $\lambda <1100$ \AA. In particular, the
${\rm C^+}$/C ratio is increased and the [\ion{C}{i}] intensity decreased if 
the UV
intensity drops less steeply at the shortest wavelengths. In that
case, the CO self-shielding can be maintained with an only slightly
higher gas mass while the relative importance of C photoionization to
CH photodissociation is decreased, leading to a lower C abundance.

\subsection{Gas/dust ratio}
The spatial dust distribution considered in this paper has a total mass of 
$\sim 2.2\,M_\oplus$ for grains with sizes $1\,{\rm \mu m}<a<1\,{\rm cm}$. 
Together with the gas mass of 80 $M_\oplus$ this would give an overall gas/dust
mass ratio of 36, which cannot be distinguished from the interstellar value
of 100 given the uncertainties in the gas mass. This may indicate that gas and 
dust are dissipated on similar timescales. However,
the upper limit of the grain size distribution is not well constrained, and the
total dust mass depends strongly on this parameter. Furthermore, it is 
uncertain what part (if any) of the dust population is debris (i.e. the product
of collisions between meter-sized bodies) and what part is ``first generation''
aggregates. Thus the derived gas/dust ratio has only limited meaning.

\subsection{Second generation gas?}

There is a possibility that the gas in an evolved disk such as that 
around HD~141569A, is entirely hydrogen-poor, second generation gas, i.e. gas
originating from evaporating solid bodies instead of the original interstellar 
cloud. The CO photodissociation timescale is short, $\sim 10^3$ years (see $\S$
\ref{chemsect}). In order to reproduce the observed CO mass of 
$4\times 10^{-4}\,M_\oplus$, CO has to evaporate from comets (or similar solid
bodies) at a rate of $4\times 10^{-7}\,M_\oplus\,{\rm yr^{-1}}$. If one takes a
mass of $10^{18}$~g ($2\times 10^{-10}\,M_\oplus$) and a CO mass fraction of 
10\% as typical numbers for solar system comets \citep[see][]{delsem85} this 
translates into a required evaporation rate of $2\times 10^4$ comets per year, 
which seems too high to be realistic.

\section{Conclusions}
A model of the HD~141569A disk has been constructed which calculates the 
chemistry and gas temperature in a quasi 2-dimensional formalism and predicts
the resulting emission lines. The main conclusions are as follows:
\begin{itemize}
\item A gas distribution with a total mass of $80\,M_\oplus$ starting at 80 
AU from the central star produces the best fit to the observed CO emission 
lines for this object. Further observations of gas-phase lines will be needed 
to corroborate this result. The gas distribution can be tested through 
spatially resolved CO observations. The [\ion{C}{i}] fine-structure lines are 
good tracers of gas 
mass, but the [\ion{C}{ii}] fine-structure line shows only a very weak 
variation with disk mass.
\item The CO chemistry of the outer disk is very sensitive to the total
gas mass due to the coupled effects of high densities (which drives carbon to 
its molecular form) and shielding of the stellar and interstellar radiation 
fields.
\item The interstellar radiation field plays a significant role in the 
dissociation of ${\rm H_2}$ and CO, because the lines through which these 
molecules dissociate become optically thick toward the central star.
\item PAHs are found to be very important in the chemistry. Because the grains 
in this disk are large, the most efficient way to form ${\rm H_2}$ is via PAHs,
and therefore they affect the formation of all molecules.
\item Carbon is found to be predominantly in neutral atomic form due a combination of efficient self-shielding of carbon photoionization and rapid photodissociation of chemical precursors of CO by the stellar radiation field. This conclusion depends strongly on the shape of the stellar radiation field at $\lambda < 1100\,$\AA.
\item The current model reproduces the CO $J=3\to 2$ line profile observed by 
\citet{dent05} quite well. The sharp peak at the red side of the spectrum is 
not reproduced due to the cylindrical symmetry inherent in the model. Also the
model uses a azimuthally averaged dust distribution while the real distribution
is noticeably sharper, which also explains the shallower profile. The current 
model ignores the possibility of a clumpy gas distribution, which would also 
result in sharper profiles. 
\end{itemize}
The calculations of the gas mass of the HD~141569 disk presented here 
indicate that transitional disks retain a fraction of their original gas
content after a few Myr. 
The remaining gas is not enough to form giant 
planets, but mechanisms exist for planets to form on shorter timescales than 
the age of the disk \citep[e.g.][]{pollac96}. It is expected that the remaining
gas has a strong influence on the dust dynamics in the disk.

\begin{acknowledgements}

The authors are grateful to Xander Tielens for helpful discussions of the 
relevant heating rates, to David Mouillet for the deprojected HST image in 
Figure \ref{hddust}, and to Bill Dent for communicating his observational 
data in Figure \ref{spect}. They thank Michiel Hogerheijde and Floris van der 
Tak for the use of their 2-D radiative transfer code. This work was supported 
by a Spinoza grant from the Netherlands Organisation for Scientific Research 
and by the European Community's Human Potential Programme under contract 
HPRN-CT-2002-00308, PLANETS.

\end{acknowledgements}

\bibliographystyle{aa}
\bibliography{4769.bib}

\end{document}